\documentclass[conference]{IEEEtran}
\usepackage{graphicx}
\usepackage{amssymb}
\usepackage{cite}

\ifCLASSINFOpdf
\else
\fi
\usepackage{amsmath}

\begin{document}

\title{A CT image based finite element modelling to predict the mechanical behaviour of human arm}

\author{\IEEEauthorblockN{Amit Kumar Bedaka}
\IEEEauthorblockA{Department of Mechanical Engineering,\\Indian Institute of Information Technology Design\\ and Manufacturing
Kancheepuram, Chennai-600 127,\\ Tamilnadu, India.\\
Email: akbedaka@gmail.com}
\and
\IEEEauthorblockN{Ponnusamy Pandithevan}
\IEEEauthorblockA{Department of Mechanical Engineering,\\Indian Institute of Information Technology Design\\ and Manufacturing
Kancheepuram, Chennai-600 127,\\ Tamilnadu, India.\\
Email: ppthevan@iiitdm.ac.in}}


%


\maketitle

\begin{abstract}
In the present work, complex irregular bones and joints of the complete human arm were developed in a computer aided design environment. Finite element analysis of an actual human arm was done to identify the distribution of stress using von-Mises stress and maximum principal stress measures. The results obtained from the present study revealed the region where, maximum stress was developed for different loading and boundary conditions with different joint rotations as obtained in the actual human arm. This subject specific analysis helps to analyse the region of the arm in which the risk is more.
\end{abstract}


%
\IEEEpeerreviewmaketitle

\section{Introduction}

The human arm poses exemplary characteristics to
perform multi-tasks simultaneously. The arm poses soft
and hard tissues to act in both elastic and plastic manner
to handle soft, tender material with care and carry robust
loads. The structure is rigid enough to take shock loads
and impact loads. The load capacity of human arm, though studied through experiments cannot be universally
applicable to all because of variation of properties and
micro structure. At this juncture, in order to replace or
repair human arm, a thorough study on stress and strain
analysis is required. A finite element study of several
human body parts is done in literature.

A human arm is having complex joints and capable of
performing various moments. The pronation/supination
and flexion/extension moment arm magnitudes with
elbow angle were represented well by the computer
model \cite{Mur95}. The trajectory of the human arm is traced \cite{Abe82}
in order to investigate the strategies used to plan and 
control, multi joint arm trajectories. Zadpoor have studied the vibrations of the humerus bone by
attaching two hemispheres to its both ends\cite{Zad06}. The bone is
considered as a linear-elastic, isotropic and homogeneous
material. The mechanical properties of the human arm are
modulated during the examination of isometric force
regulation tasks \cite{Per04}.

The apparent elastic modulus and viscous behavior of
cardiac and skeletal muscle and vascular endothelium
would differ due to functional and structural differences\cite{Mat01}. Analyses of the mechanics of narrowly contained
tissues have used Poisson's ratios to study the significant
effect on the results\cite{Van93}. Mechanical behavior of $42$ fresh
human cadaver lumbar motion segments in flexion,
extension, lateral bending and torsion is examined for
different loading conditions\cite{Sch79}. Finite element analysis
described an easy method to simulate the kinetics of
multi-body mechanisms\cite{Van94}. It is used in order to develop
a musculo-skeletal model of the shoulder mechanism.
Finite element model of the human hand-arm system is
used to derive natural frequencies and mode shapes\cite{Ade14}.
In the present work, in order to identify the weaker region
of the human arms during movement, various case studies
were conducted. In addition, an image based computer model of the human arm with soft tissue, bone and bone
marrow was developed to perform the analysis more
accurately.

\section{THREE DIMENSIONAL CAD MODELING OF HUMAN ARM}

\begin{figure}[!t]
\centering
\includegraphics[width=2.9in]{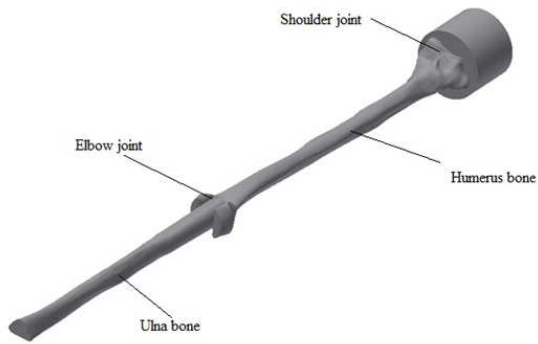}
\caption{Sub-assembly of human arm.}
\label{fig_1}
\end{figure}

\begin{figure}[!t]
\centering
\includegraphics[width=2.5in]{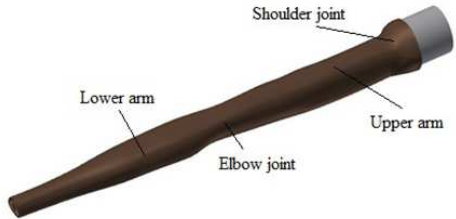}
\caption{Complete assembly of the soft tissue, bone 
and bone marrow.}
\label{fig_2}
\end{figure}

Effective methods for the conversion of computed
tomography data into computer-aided design models are
still in development\cite{You12}. In the present work, the
following three different process paths for the generation
of a computer-aided design model was used. The human arm contains different types of bones and
each bone unique in its geometry. Coming from the
shoulder the first is humerus head, which has most complex 
geometry. It is located at the shoulder joint
connecting the shoulder joint and the humerus. Humerus
is the midsection of the arm lies between the head and
elbow. It is comparatively less complex than humerus
head, but it is the longest part and maximum load taking
bone. The ulna is the last bone connecting the wrist and
the elbow, as shown in Fig. \ref{fig_1}. The complete modeling to the human arm with soft tissue,
bone and bone marrow have been performed in Autodesk Inventor$^\circledR$, 
as shown in Fig. \ref{fig_2}.

\begin{figure}[!t]
\centering
\includegraphics[width=1.8in]{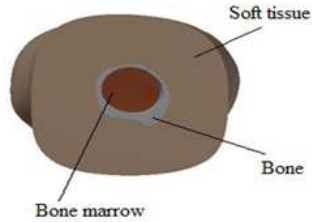}
\caption{Cross sectional view of the complete assembly.}
\label{fig_3}
\end{figure}

\begin{figure}[!t]
\centering
\includegraphics[width=3.4in]{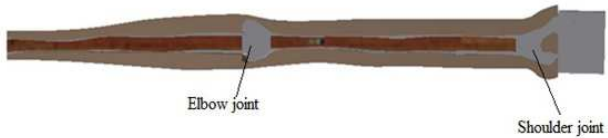}
\caption{Longitudinal section of the complete assembly.}
\label{fig_4}
\end{figure}

Cross sectional view, as shown in Fig. \ref{fig_3} has the three
different parts of human arm, where outer part is soft
tissue followed by the bone and the inner part is bone
marrow. Fig. \ref{fig_4} shows the shoulder joint with spherical joint to provide 3 DOF and elbow joint with one DOF as like
hinge joint. Finally, the complete assembly is exported in
the IGES format to do the further finite element analysis.

The CAD model was imported into ANSYS workbench. Subsequently, the material properties of the soft tissue, bone and bone marrow were keyed into the workbench. Material properties were assigned by selecting individual part of the model. The type of contacts and joints were assigned between them. Prior to connections, co-ordinates were defined using the global co-ordinate system. Different mesh size was attributed and convergence study was performed using linear tetrahedral element. In static structure, the loading and boundary conditions were input. Finally, the stress output required from the simulation results was specified to locate the maximum stress region in the solution.

\section{RESULTS AND DISCUSSION}
Analysis of the individual bone and complete assembly
is performed in the ANSYS workbench to perform static
structural analysis. To calculate the effects of steady
loading conditions on the individual parts as well on the
complete assembly by using the boundary condition as
like human arm motion. In static analysis the material
properties of bone, bone marrow and soft tissue is
assigned according to the values found by [3], [5], [6], [9].
Similarly, the loading and boundary conditions for
different movements of the human arm is applied in our work as used by [1], [2], [11].

\subsection{Static analysis of humerus and ulna bone assembly (Stress):}
\begin{figure}[!t]
\centering
\includegraphics[width=3.0in]{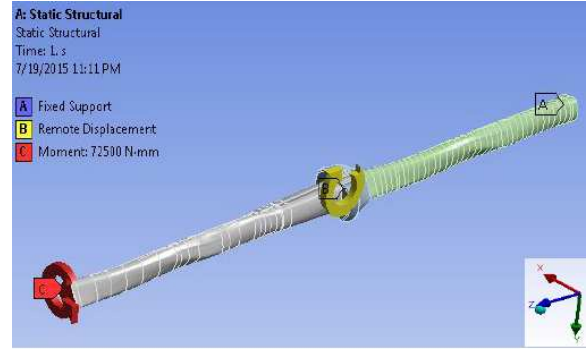}
\caption{Boundary condition for the humerus and ulna bone assembly.}
\label{fig_5}
\end{figure}

In this case, the humerus and ulna bone assembled by
elbow joint for the given flexion movement with the
boundary condition, as shown in Table \ref{table_1} with respect to the co-ordinate, as shown in Fig. \ref{fig_5}. The shoulder joint is fixed and the elbow joint is rotated along the $x-$axis, as shown in Fig. \ref{fig_5}. The applied moment is $72500Nmm$\cite{Ask87} along $z-$axis at the tip of the ulna bone, as shown in Fig. \ref{fig_5}. The humerus and ulna bone assembly is meshed using tetrahedral mesh type with $1.25mm$ mesh size.

\begin{table}[!t]
\caption{Boundary condition of Flexion movement of elbow 
joint with fixed shoulder joint.}
\label{table_1}
\centering
\begin{tabular}{|c|c|c|}
\hline
{\bf Serial No.} & {\bf Co-ordinate} & {\bf Degree}\\
\hline
1 & $R_x$ & 145$^\circ$\\
\hline
2 & $R_y$ & 0$^\circ$\\
\hline
3 & $R_z$ & 0$^\circ$\\
\hline
\end{tabular}
\end{table}

\subsubsection{Case I}
Flexion movement of ulna bone (elbow joint)
is achieved with the given boundary condition as, shown
in Table \ref{table_1} w.r.t the co-ordinates, as shown in Fig. \ref{fig_5}.
The shoulder joint is fixed and the elbow joint is rotated
along the $x-$axis, as shown in Fig. \ref{fig_5}. The applied moment
is $72500Nmm$\cite{Ask87} at the tip of the ulna bone, as shown in
Fig. \ref{fig_5}. For the given boundary and loading condition
static analysis of the ulna and humerus bone is performed
to study the different stress, as shown in the Table \ref{table_2}.

\begin{table}[!t]
\caption{Different stress of Flexion movement of elbow joint 
with fixed shoulder joint.}
\label{table_2}
\centering
\begin{tabular}{|c|c|c|}
\hline
{\bf Serial No.} & {\bf Stress} & $ {\bf MPa}$\\
\hline
1 & Equivalent stress & 224.67\\
  & (von-Mises stress) & \\
\hline
2 & Maximum Principal stress & 157.18\\
\hline
3 & Max Shear stress & 129.14\\
\hline
4 & Normal stress & 102.82\\
\hline
\end{tabular}
\end{table}

\subsubsection{Case II}
Extension movement of ulna bone (elbow
joint) is achieved with the given boundary condition, as
shown in Table \ref{table_3} w.r.t the co-ordinates, as shown in Fig. \ref{fig_5}. The shoulder joint is fixed and the elbow joint is also
constrained in all the three axes of rotation. The applied
moment is $42100Nmm$\cite{Ask87} along $z-$axis at the tip of the
ulna bone. For the given boundary and loading condition
static analysis of the ulna and humerus bone is performed
to study the different stress, as shown in the Table \ref{table_4}.

\begin{table}[!t]
\caption{Boundary condition of Extension movement of elbow 
joint with fixed shoulder joint.}
\label{table_3}
\centering
\begin{tabular}{|c|c|c|}
\hline
{\bf Serial No.} & {\bf Co-ordinate} & {\bf Degree}\\
\hline
1 & $R_x$ & 0$^\circ$\\
\hline
2 & $R_y$ & 0$^\circ$\\
\hline
3 & $R_z$ & 0$^\circ$\\
\hline
\end{tabular}
\end{table}

\begin{table}[!t]
\caption{Different stress of Extension movement of elbow 
joint with fixed shoulder joint.}
\label{table_4}
\centering
\begin{tabular}{|c|c|c|}
\hline
{\bf Serial No.} & {\bf Stress} & $ {\bf MPa}$\\
\hline
1 & Equivalent stress & 130.46\\
  & (von-Mises stress) & \\
\hline
2 & Maximum Principal stress & 91.273\\
\hline
3 & Max Shear stress & 74.988\\
\hline
4 & Normal stress & 59.706\\
\hline
\end{tabular}
\end{table}

The static analysis of the humerus and ulna bone is
performed to study the behavior under static loading
condition, before going for static analysis of the complete
human arm with soft tissue, humerus bone, ulna bone and
bone marrow. In the above, cases isotropic material
property of the bone is assigned and the maximum stress
is obtained near the elbow joint.

\subsection{Static analysis of muscle, bone marrow, humerus and ulna bone assembly (Stress):}
\subsubsection{Case I}
\begin{figure}[!t]
\centering
\includegraphics[width=3.0in]{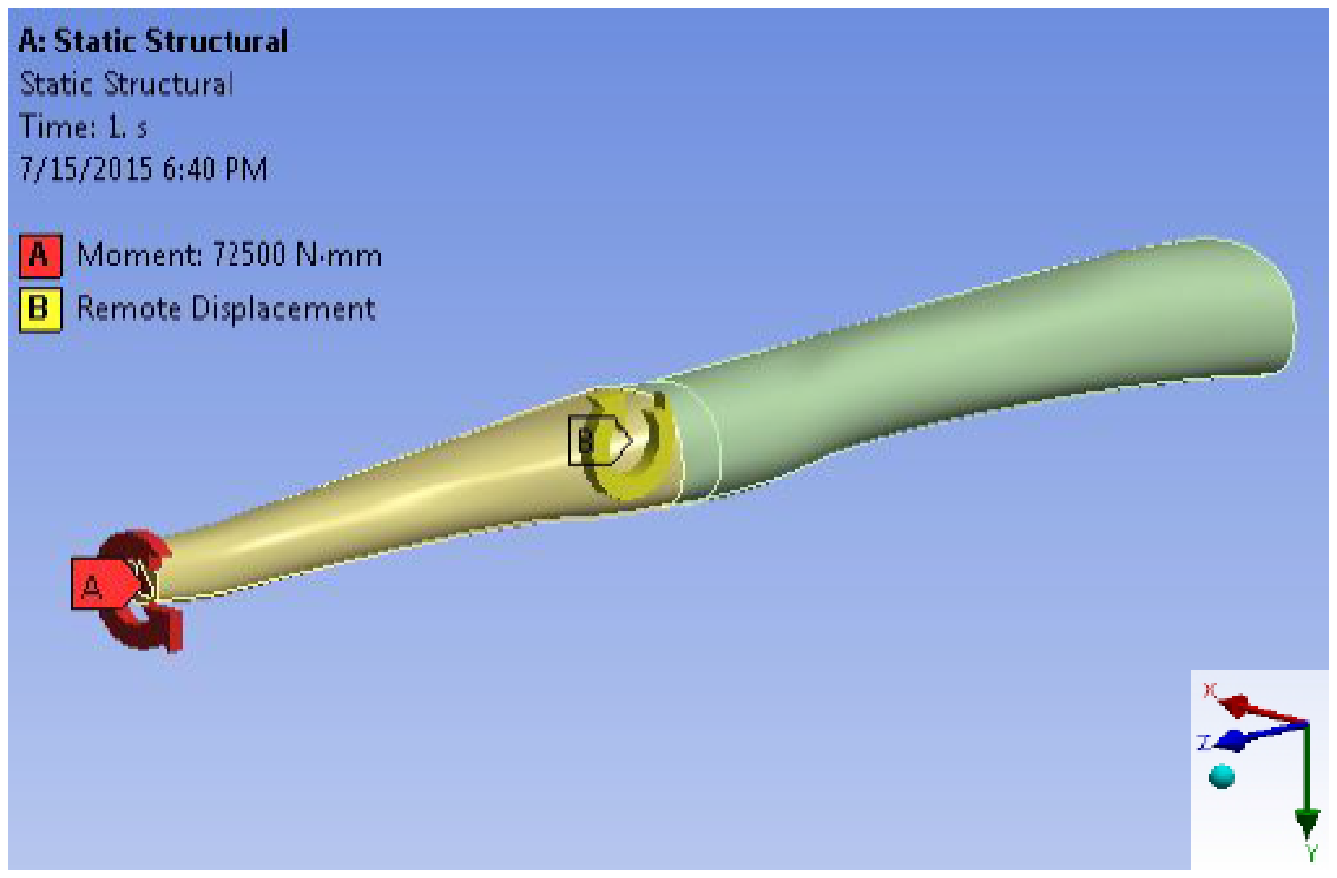}
\caption{Boundary condition for the humerus and ulna bone with soft
tissue and bone marrow.}
\label{fig_6}
\end{figure}

\begin{figure}[!t]
\centering
\includegraphics[width=3.0in]{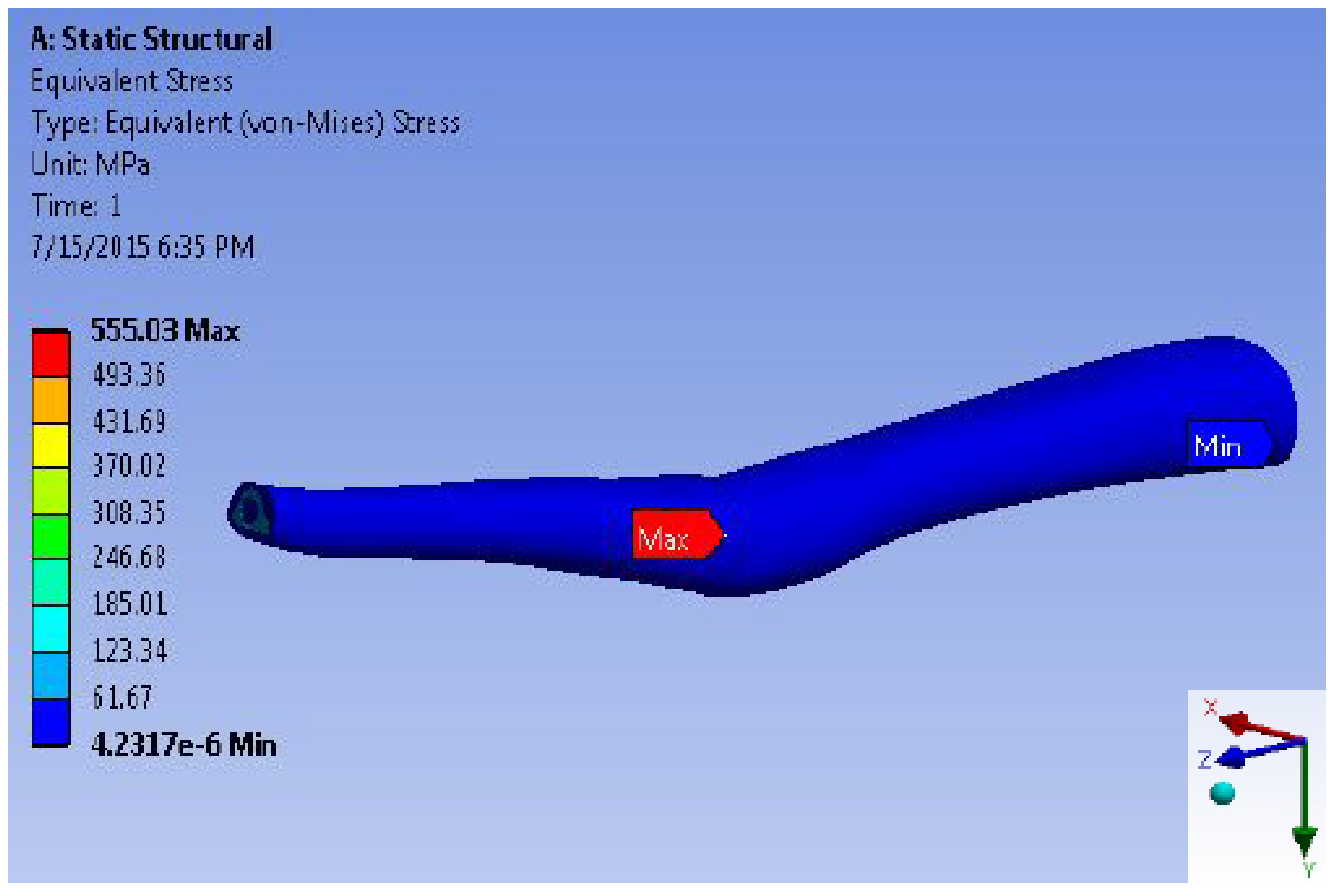}
\caption{Von-Mises stress of flexion movement with the maximum stress
location.}
\label{fig_7}
\end{figure}

Flexion movement of the elbow (elbow joint) is achieved with the given boundary condition as shown in Table \ref{table_5} w.r.t the co-ordinates, as shown in Fig. \ref{fig_6}. The shoulder joint is
fixed and the elbow joint is rotated along the $x-$axis, as shown
in Fig. \ref{fig_6}. The applied moment is $72500Nmm$\cite{Ask87} along $z-$axis
at the tip of the ulna bone, as shown in Fig. \ref{fig_6}. The
humerus and ulna bone assembly is meshed using
tetrahedral mesh type with $2.0mm$ mesh size. The complete human arm consists of soft tissue,
humerus bone, ulna bone and bone marrow. The material
properties are individually mapped to the human arm. For
the given boundary and loading condition static analysis
of the complete human arm is performed to study the
different stress, as shown in the Table \ref{table_6}. The stress is
maximum at ulna bone near to the elbow joint, as shown
in the Fig. \ref{fig_7}.

\begin{table}[!t]
\caption{Boundary condition of Flexion movement of elbow 
joint with fixed shoulder joint.}
\label{table_5}
\centering
\begin{tabular}{|c|c|c|}
\hline
{\bf Serial No.} & {\bf Co-ordinate} & {\bf Degree}\\
\hline
1 & $R_x$ & 145$^\circ$\\
\hline
2 & $R_y$ & 0$^\circ$\\
\hline
3 & $R_z$ & 0$^\circ$\\
\hline
\end{tabular}
\end{table}

\begin{table}[!t]
\caption{Different stress of Flexion movement of ulna 
joint with fixed shoulder joint.}
\label{table_6}
\centering
\begin{tabular}{|c|c|c|}
\hline
{\bf Serial No.} & {\bf Stress} & $ {\bf MPa}$\\
\hline
1 & Equivalent stress & 555.03\\
  & (von-Mises stress) & \\
\hline
2 & Maximum Principal stress & 408.51\\
\hline
3 & Max Shear stress & 320.40\\
\hline
4 & Normal stress & 150.44\\
\hline
\end{tabular}
\end{table}

\subsubsection{Case II}
\begin{figure}[!t]
\centering
\includegraphics[width=3.0in]{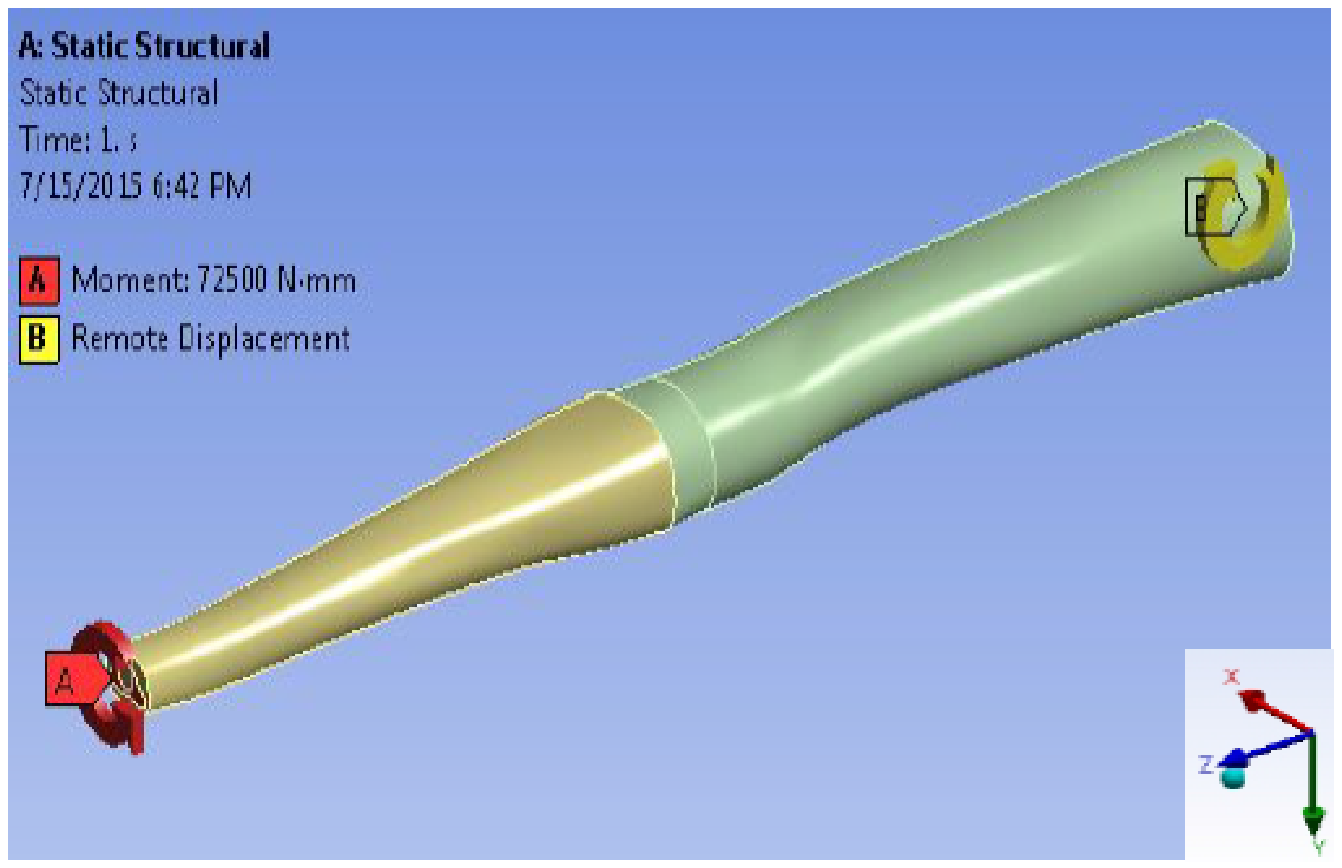}
\caption{Boundary condition for the humerus and ulna bone with soft
tissue and bone marrow.}
\label{fig_8}
\end{figure}

\begin{figure}[!t]
\centering
\includegraphics[width=3.0in]{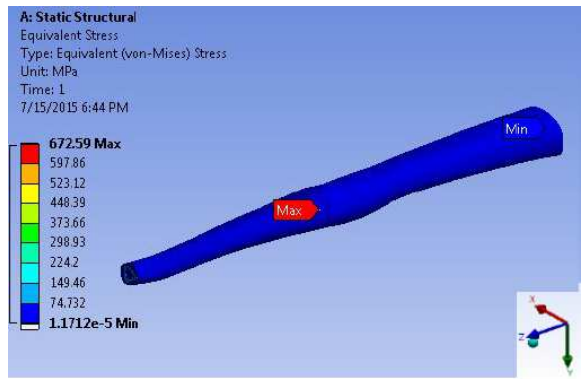}
\caption{von-Mises stress of flexion movement with the maximum 
stress location.}
\label{fig_9}
\end{figure}

Flexion movement of the shoulder (shoulder joint) is achieved
with the given boundary condition as shown in Table \ref{table_7}
w.r.t the co-ordinates, as shown in Fig. \ref{fig_8}. The elbow joint is
fixed and the shoulder joint is rotated along the $x-$axis, as shown 
in Fig. \ref{fig_8}. The applied moment is $72500Nmm$ [11] along $z-$axis
at the tip of the ulna bone, as shown in Fig. \ref{fig_8}. The
humerus and ulna bone assembly is meshed using
tetrahedral mesh type with $2.0mm$ mesh size. For the given boundary and loading condition static
analysis of the complete human arm is performed to study
the different stress which is shown in the Table \ref{table_8}. The
material properties are individually mapped to soft tissue,
humerus bone, ulna bone and bone marrow. The stress is
maximum at ulna bone near the elbow joint, as shown in
the Fig. \ref{fig_9}.

\begin{table}[!t]
\caption{Boundary condition of Flexion movement of shoulder joint.}
\label{table_7}
\centering
\begin{tabular}{|c|c|c|}
\hline
{\bf Serial No.} & {\bf Co-ordinate} & {\bf Degree}\\
\hline
1 & $R_x$ & 90$^\circ$\\
\hline
2 & $R_y$ & 0$^\circ$\\
\hline
3 & $R_z$ & 0$^\circ$\\
\hline
\end{tabular}
\end{table}

\begin{table}[!t]
\caption{Different stress of Flexion movement of shoulder joint.}
\label{table_8}
\centering
\begin{tabular}{|c|c|c|}
\hline
{\bf Serial No.} & {\bf Stress} & $ {\bf MPa}$\\
\hline
1 & Equivalent stress & 672.59\\
  & (von-Mises stress) & \\
\hline
2 & Maximum Principal stress & 697.41\\
\hline
3 & Max Shear stress & 376.88\\
\hline
4 & Normal stress & 163.91\\
\hline
\end{tabular}
\end{table}

Static structural analysis of the individual bone is
performed to understand various motions for the different
loading and boundary conditions. To identify the failure
region with complete assembly of human arm, four
different theories have been used because it's difficult to predict
the ductile and brittle nature of the bone. The measure of von-Mises stress and maximum principal stress are suitable for both brittle and ductile materials. Both the theory shows the same
maximum stress region for both assemblies.

\section{Conclusion}
The current work presents a method to model and
analyse the behavior of human arm under static loading
conditions with the assembly of the humerus and ulna
bone. In this work, complex and irregular bones and
joints of the human arm were modeled in computer-aided
design environment. Finite element analysis of actual
human arm was done by assigning the individual material
properties to identify maximum stress region using von-
Mises stress, maximum principal stress and maximum
shear stress because it’s difficult to predict the
ductile and brittle nature of bone. This study can help to
reduce the failure at maximum stress region by taking
appropriate care. Maximum stress is also obtained to
identify the week region, where the stress has maximum
values. The study can be extended to predict the behavior
in different movements of the human arm.




%

\end{document}